\magnification=\magstep 1
\tolerance 500
\rightline{IASSNS-HEP-96/39}
\vskip 3 true cm
\centerline{\bf Algebra of Conserved Generators and Statistical Ensembles}
\centerline{\bf in}
\centerline{{\bf Generalized Quantum Dynamics} \footnote{*}{Presented as
a talk by the second author at the Workshop on Algebraic Approaches to
Quantum Dynamics, May 7-12, 1996, at the Fields Institute for Research
in Mathematical Sciences, Toronto, Ontario, Canada.}}
\vskip 1 true cm
\centerline{Stephen L.  Adler and L.P. Horwitz\footnote{**}{On sabbatical leave from 
School of Physics and Astronomy, Raymond and Beverly Sackler Faculty
of Exact Sciences, Tel Aviv University, Ramat Aviv, Israel, and
Department of Physics, Bar Ilan University, Ramat Gan, Israel.}}
\smallskip
\centerline{School of Natural Sciences, Institute for Advanced Study}
\centerline{Princeton, N.J. 08540}
\vskip 2 true cm
\noindent
{\it Abstract \/}: We study here the algebraic structure of the
conserved generators from which the microcanonical and canonical
ensembles are constructed on an underlying generalized quantum
dynamics, and the flows they induce on the phase space. We
also discuss briefly the structure of the  microcanonical and
canonical ensembles.   

\vfill
\eject
\noindent 
{\bf 1. Introduction}
\smallskip
\par It has recently been shown$^1$, by application of
statistical mechanical methods to determine the canonical ensemble
governing the equilibrium distribution of operator initial values, 
 that complex quantum field theory can emerge as a statistical 
approximation to an underlying generalized quantum dynamics.$^{2,3}$ This
result was obtained by an argument based on a Ward identity analogous to
the equipartition theorem  of classical statistical mechanics.$^1$  In
the following, we describe some features of generalized quantum
dynamics and the motivation for its development. We then describe how
the equilibrium ensembles are constructed, and briefly review the
emergence of the usual complex quantum field theory in the ensemble
average.
\par The field variables associated with a hypercomplex quantum
theory, such as quaternionic quantum mechanics, form an essentially
non-commutative set of functions, even when they are not
second-quantized. The construction of a dynamical theory then requires
a generalized structure, which is the main subject of our discussion
here. To understand why this is so,
consider the Heisenberg picture of a scalar field.$^3$  The generator of
evolution in time is taken as an anti-self-adjoint
operator, since there is, in general,  no natural complex unit to extract a
self-adjoint operator from the representation of a one-parameter
unitary group. The Heisenberg picture for a quaternionic field of the
form
$$ \phi(x) = \sum_{A=0}^3 \phi_A e_A, \eqno(1.1)$$
where $e_0 =1$ and $e_1,\,e_2,\,e_3$ are the quaternionic units
satisfying the cyclic relations
$$ e_1e_2 = e_3, \qquad e_A^2 = -1 \qquad A=1,2,3, \eqno(1.2)$$
is given by
$$\eqalign{\phi_H(x,t)& = e^{{\tilde H}t} \phi(x)e^{-{\tilde H}t} \cr
&= \phi_0(x,t) + \sum_{A=1}^3 \phi_{AH}(x,t) e_A(t),\cr}
\eqno(1.3)$$
where the $\{\phi_{AH}(x,t)\}$ are kinematically independent, i.e.,
the commutation relations are the same as the Schr\"odinger picture
fields. However, if we expand over the original set of quaternionic
units $\{e_A\}$ on the components defined by
$$ \phi_H(x,t) = \sum_{A=0}^3 \phi_H(x,t)_A e_A, \eqno(1.4)$$
then the $\{\phi_H(x,t)_A \}$ are {\it not} kinematically independent.
As an illustrative example, consider the simple case in which ${\tilde
H}$ is some real valued operator multiplied by $e_1$. Since
$$\eqalign{ e_{1H}(t) &= e_1\cr
            e_{2H}(t) &= e^{{\tilde H}t} e_2 e^{-{\tilde H} t}=
e^{2{\tilde H}t} e_2 \cr
            e_{3H}(t) &= e^{{\tilde H}t} e_3 e^{-{\tilde H}t} =
e^{2{\tilde H}t}e_3, \cr } \eqno(1.5)$$
the coefficients of the constant $\{e_A\}$ are
$$\eqalign{ \phi_H(x,t)_1 &= \phi_{1H}(x,t) \cr
            \phi_H(x,t)_2 &= \phi_{2H}(x,t)e^{2{\tilde H}t} \cr
            \phi_H(x,t)_3 &= \phi_{3H}(x,t)e^{2{\tilde H}t}. \cr}
\eqno(1.6)$$ 
Due to the presence of the additional factor containing ${\tilde H}$,
the commutation relations are not the same as those of the Schr\"odinger
fields, and, in fact,
$$ [\phi_H(x,t)_A, \, \phi_H(x',t)_B ] \neq 0. \eqno(1.7)$$
The same result is true for gauge fields of the form $B_\mu = \sum
B_{\mu A} e_A$. A similar structure was found in the studies of ref.
4, a program concerned with the static limit of generalized non-Abelian gauge
theories. In order to construct a c-number action for theories of this
type, the notion of a {\it total trace } was introduced$^{2,3}$, where
both the quantum state and algebraic representation indices are summed
over in the trace.  As we shall see in the next section, one can define a total
trace Lagrangian and Hamiltonian over an underlying phase space of
quantum fields which results in a symplectic dynamics with  many of
the properties of classical mechanics.  We shall the describe the
flows induced by an important class of conserved operators, and show
how a statistical mechanics can be constructed on the phase space of
the underlying quantum fields.
\bigskip
\noindent{\bf 2. Generalized quantum dynamics}
\smallskip
\par Generalized quantum dynamics$^{2,3}$ is an analytic mechanics on
a symplectic set of operator valued variables, forming an operator
valued phase space $\cal S$.  These variables are defined as the set of
linear transformations\footnote{$^\dagger$}{In general, local
(noncommuting) quantum fields.}
 on an underlying real, complex, or quaternionic
Hilbert space (Hilbert module), for which the postulates of a real,
complex, or quaternionic quantum mechanics are satisfied$^{3,5-8}$.  The
dynamical (generalized Heisenberg) evolution, or flow, of this phase space is
generated by the total trace Hamiltonian ${\bf H} = {\bf Tr} H$, where
$$ \eqalign{{\bf Tr} {\rm O} &= {\rm Re Tr} (-1)^F {\rm O} \cr
                              &= {\rm Re} \sum_n \langle n \vert (-1)^F
{\rm O} \vert n \rangle,\cr} \eqno(2.1)$$   
$H$ is a function of the operators $\{ q_r(t) \}, \{p_r(t) \},\ \ {\rm
r}= 1,2, \dots,N$ (realized as a sum of monomials, or a limit of a
sequence of such sums; in the general case of local noncommuting
fields, the index $r$ contains continuous variables), and $(-1)^F$ is a grading operator with
eigenvalue $1(-1)$ for states in the boson (fermion) sector of the
Hilbert space. Operators are called bosonic or fermionic in type if they
commute or anticommute, respectively, with $(-1)^F$; for each $r$, $p_r$ 
and $q_r$ are of the same type.
\par The derivative of a total trace functional with respect to some
operator variation is defined with the help of the cyclic property of
the ${\bf Tr}$ operation.  The variation of any monomial ${\rm O}$
consists of terms of the form ${\rm O}_L \delta x_r {\rm O}_R$, for
$x_r$ one of the $\{q_r \}, \{p_r\}$,
which, under the ${\bf Tr}$ operation, can be brought to the form
$$ \delta {\rm{\bf O}} = \delta {\bf Tr}{\rm O} = \pm {\bf Tr}
{\rm O}_R {\rm O}_L \delta x_r, $$
so that sums and limits of sums of such monomials permit the
construction of 
$$ \delta {\rm{\bf O}} = {\bf Tr} \sum_r {\delta {\rm{\bf O}} \over
\delta x_r } \delta x_r,\eqno(2.2) $$
uniquely defining ${\delta {\rm{\bf O}}}/{\delta x_r}$.
\par Assuming the existence of a total trace Lagrangian$^{2,3}$ ${\bf
L} = {\bf L} (\{q_r \}, \{ {\dot q}_r \} )$, the variation of the
total trace action
$$ {\bf S} = \int_{-\infty}^\infty  {\bf L} (\{q_r \}, \{ {\dot q}_r
\}) dt   \eqno(2.3) $$
results in the operator Euler-Lagrange equations
$$  {\delta {\bf L} \over {\delta q_r}} - {d \over {dt}} {{\delta {\bf
L}} \over {\delta \dot q_r} } = 0 . \eqno(2.4)$$
As in classical mechanics, the total trace Hamiltonian is defined as a  
Legendre transform, 
$$ {\bf H} = {\bf Tr} \sum_r p_r \dot q_r - {\bf L},\eqno(2.5) $$
where 
$$ p_r = { {\delta {\bf L}} \over {\delta \dot q_r}}.  \eqno(2.6)$$ 
 It then follows from $(2.4)$ that 
$${ {\delta {\bf H}} \over {\delta q_r} } = -\dot p_r  \qquad { {\delta
{\bf H}} \over {\delta p_r}} = \epsilon_r \dot q_r,\eqno(2.7) $$
where $\epsilon_r = 1(-1)$ according to whether $p_r, q_r$ are of 
bosonic (fermionic) type.  
\par Defining the generalized Poisson bracket
$$ \{{\bf A}, {\bf B} \} = {\bf Tr} \sum_r \epsilon_r \left( {\delta
{\bf A} \over \delta q_r} {\delta {\bf B}\over \delta p_r} -
{\delta {\bf B} \over \delta q_r} {\delta {\bf A}\over \delta
p_r}\right), \eqno(2.8a) $$
one sees that
$$ {d{\bf A} \over dt } = {\partial {\bf A} \over \partial t} +
\{{\bf A}, {\bf H} \}. \eqno(2.8b) $$
Conversely, if we define 
$$ {\bf x}_s(\eta) = {\bf Tr} (\eta x_s) , \eqno(2.9a)$$
for $\eta$ an arbitrary, constant operator (of the same type as $x_s$, 
which denotes here $q_s$ or $p_s$), then
 $$  {d{\bf x}_s(\eta)\over dt} = {\bf Tr} \sum_r \epsilon_r \left( {\delta
{\bf x}_s(\eta) \over \delta q_r} {\delta {\bf H}\over \delta p_r} -
{\delta {\bf H} \over \delta q_r} {\delta {\bf x}_s(\eta)\over \delta
p_r}\right), \eqno(2.9b)$$
and comparing the coefficients of $\eta$
on both sides, one obtains the Hamilton equations $(2.7)$ as a
consequence of the Poisson bracket relation $(2.8b)$.
\par The Jacobi identity is satisfied by the Poisson bracket of
$(2.8a)$,$^9$ and hence the total trace functionals have many of the
properties of the corresponding quantities in classical mechanics.$^{10}$
In particular, canonical transformations take the form 
$$\delta {\bf x}_s(\eta)=\{ {\bf x}_s(\eta), {\bf G} \}, \eqno(2.10a)$$
which implies that 
$$\delta p_r= -{\delta {\bf G} \over \delta q_r}~,~~~
\delta q_r=\epsilon_r  {\delta {\bf G} \over \delta p_r}~,\eqno(2.10b)$$
with the generator ${\bf G}$ any total trace functional constructed from
the operator phase space variables.  Time evolution then corresponds to the 
special case ${\bf G}={\bf H} dt$.
\bigskip
\noindent{\bf 3. Flows and algebras associated with conserved
operators}
\smallskip
\par  The operator
$$ \eqalign{{\tilde C} &= \sum_r (\epsilon_r q_r p_r - p_r q_r) \cr
&=\sum_{r,B} [q_r,p_r] - \sum_{r,F} \{q_r,p_r\},\cr}
\eqno(3.1)$$
where the sums are over bosonic and fermionic pairs, respectively, is
conserved under the evolution $(2.7)$ induced by the total trace
Hamiltonian.  When the equations of motion
induced by the  Lagrangian ${\bf L}$ coincide with those induced by
the ungraded total trace of the same Lagrangian (there are many models
which have this property$^1$),
$$ {\hat {\bf L}} = {\rm Re Tr} L, $$
without the factor $(-1)^F$, the corresponding ungraded total trace
Hamiltonian ${\hat {\bf H}}$ is conserved; it may therefore be
included as a constraint functional in the canonical ensemble, along
with the new conserved operator
$$ \eqalign{{\hat{\tilde C}} &= \sum_r [q_r, p_r]\cr
 &= \sum_{r,B} [q_r,p_r] + \sum_{r,F} [q_r,p_r] .\cr}
\eqno(3.2) $$
\par  We now study the action of general total trace 
functionals projected from $\tilde C$ and $\hat{\tilde C}$  
as generators of canonical transformations on the phase space. 
 We first remark that it was pointed out in ref. 1 that a canonical
generator of unitary transformations on the basis of the underlying
Hilbert space has the form
$$ {\bf G_{\tilde f}} = - {\bf Tr} \sum_r [\tilde f, p_r] q_r, \eqno(3.3) $$
where ${\tilde f}$ is bosonic.
Using $(3.1)$ and the cyclic properties of ${\bf Tr}$, one sees that
$$\eqalign{{\bf G_{\tilde f}} &= - {\bf Tr}{\tilde f} \sum_r  (p_r q_r-
\epsilon_r q_r p_r) \cr
 &= {\bf Tr} {\tilde f}{\tilde C}. \cr} \eqno(3.4)$$
We thus see that the conserved operator $\tilde C$ has the additional 
role of inducing the action of unitary transformations
on the underlying Hilbert space. 

\par That this action preserves the
algebraic properties of functionals of the type ${\bf G}_{\tilde f}$  can be seen
by computing the Poisson bracket
$$ \{{\bf G}_{\tilde f}, {\bf G}_{\tilde g} \} = 
{\bf Tr}\sum_r \epsilon_r \bigl(
{\delta {\bf G}_{\tilde f} \over \delta q_r} {\delta {\bf G}_{\tilde g} \over \delta
p_r} - {\delta {\bf G}_{\tilde g} \over \delta q_r} 
{\delta {\bf G}_{\tilde f} \over \delta
p_r} \bigr). \eqno(3.5)$$
We use the result that
$$ \eqalign{\delta {\bf G}_{\tilde f} &= {\bf Tr} {\tilde f} \delta {\tilde C} \cr
&= {\bf Tr} \sum_r \{\epsilon_r({\tilde f} q_r -q_r {\tilde f}) \delta p_r 
-  ( {\tilde f} p_r - p_r {\tilde f}) \delta q_r\} \cr} \eqno(3.6)$$
 to obtain
$$ \eqalign{ {\delta {\bf G}_{\tilde f} \over \delta q_r} &= - [{\tilde
f}, p_r] ,
\cr
{\delta {\bf G}_{\tilde f} \over \delta p_r} &= \epsilon_r [{\tilde f}, q_r],\cr}
\eqno(3.7)$$
and hence, expanding out the  commutators, one finds that
$$ \eqalign{\{ {\bf G}_{\tilde f}, {\bf G}_{\tilde g} \} &= -{\bf Tr} \sum_r
[{\tilde f}, {\tilde g}] (p_r q_r - \epsilon_r q_r p_r) \cr
&= {\bf Tr} [{\tilde f}, {\tilde g}] {\tilde C} \cr
&={\bf G}_{[\tilde f, \tilde g]}  .\cr
} \eqno(3.8)$$
These relations, corresponding to the group properties of integrated
charges in quantum field theory, can be generalized to a ``local''
algebra.  Defining
$$ {\bf G}_{\tilde f r} = {\bf Tr} {\tilde f}{\tilde C}_r, \eqno(3.9)$$
where
$$ {\tilde C}_r = \epsilon_r q_rp_r -  p_rq_r, \eqno(3.10)$$
one obtains in the same way that
$$ \{{\bf G}_{\tilde f r}, {\bf G}_{\tilde g s}\} = \delta_{rs} {\bf G}_{
[{\tilde f}, {\tilde g}] r}. \eqno(3.11)$$

\par In studying the flows induced by conserved operators, we shall
also need the properties of generators projected from $\hat {\tilde C}$.
We therefore define
$$ {\hat{\bf G}_{\tilde f}} = {\bf Tr} {\tilde f}{\hat{\tilde
C}}.\eqno(3.12)$$  
Substituting $(3.2)$, we find that the operator derivatives of $\hat{\bf G}_
{\tilde f}$ with respect to the phase space variables are 
$$ \eqalign{ {\delta \over \delta q_r} {\hat{\bf G}_{\tilde f}}
 &=  - (-1)^F[(-1)^F \tilde f,  p_r] 
 =-(\tilde f p_r -\epsilon_r p_r \tilde f),  \cr
{\delta \over \delta p_r}{\hat {\bf G}_{\tilde f}}
&= (-1)^F [(-1)^F \tilde f, q_r]
=\tilde f q_r-\epsilon_r q_r \tilde f. \cr} \eqno(3.13)$$
Computing Poisson brackets in the same way as above, we find that the 
algebra of the generators ${\bf G}_{\tilde f}$ and 
$\hat{\bf G}_{\tilde f}$ closes, 
$$\eqalign{ 
\{ \hat{\bf G}_{\tilde f}, \hat{\bf G}_{\tilde g} \} = 
&{\bf G}_{[\tilde f, \tilde g]},\cr
\{ \hat{\bf G}_{\tilde f}, {\bf G}_{\tilde g} \}
=&\{ {\bf G}_{\tilde f}, \hat{\bf G}_{\tilde g} \}
=\hat{\bf G}_{[\tilde f, \tilde g]}, \cr
}\eqno(3.14)$$
giving a structure analogous to that of the vector and axial-vector charge 
algebra in quantum field theory.  The algebra of (3.8) and (3.14) can 
be diagonalized into two independent algebras 
$${\bf G}_{\pm \tilde f}={1\over 2} ({\bf G}_{\tilde f} \pm 
\hat{\bf G}_{\tilde f}), \eqno(3.15)$$
which obey the algebra
$$ \eqalign{
\{ {\bf G}_{\pm \tilde f} , {\bf G}_{\pm \tilde g} \}=&
{\bf G}_{\pm [\tilde f, \tilde g]}, \cr
\{ {\bf G}_{+ \tilde f}, {\bf G}_{- \tilde g} \}=&0. \cr
}\eqno(3.16)$$
Defining a ``local'' version of $\hat{\bf G}_{\tilde f}$ by 
$$\hat{\bf G}_{\tilde f r} = {\bf Tr} {\tilde f} \hat{\tilde C}_r, 
\eqno(3.17)$$
where
$$ \hat{\tilde C}_r =  q_rp_r -  p_rq_r, \eqno(3.18)$$
the algebras of $(3.14)$ and $(3.16)$ can be converted to local versions 
analogous to $(3.11)$.

\par We now turn to the flows associated with  ${\bf G}_{\tilde f}$ and 
$\hat{\bf G}_{\tilde f}$ when used as canonical generators.
Beginning with ${\bf G}_{\tilde f}$, we consider its action on the 
functional ${\bf x}_s(\eta)$ defined in $(2.9a)$, for which
$\delta {\bf x}_s(\eta)={\bf Tr} \eta \delta x_s$.  
Defining a parameter $\gamma$ along the motion generated by
${\bf G}_{\tilde f}$,  we choose $\delta x_s$ as $d x_s / {d \gamma}
$, so that by $(2.10a)$ we have
$$ d {\bf x}_s(\eta) = \{{\bf x}_s(\eta), {\bf G}_{\tilde f} \} d\gamma 
.\eqno(3.19)$$
Comparing $(2.10b)$ with $(3.5)$ and $(3.19)$ gives
$$ \eqalign{{d q_s \over {d \gamma}} &= [\tilde f, q_s], \cr
{d p_s \over {d \gamma}} &= [\tilde f,p_s]. \cr} \eqno(3.20)$$
In both the boson and fermion sectors we see that, as a solution of
the differential equations $(3.20)$, ${\bf G}_{\tilde f}$
induces the action of a unitary group generated by ${\tilde f}$,
$$ x_s(\gamma)= e^{{\tilde f}\gamma} x_s(0) e^{-{\tilde f}\gamma}.
\eqno(3.21)$$
The transformation $(3.21)$ is unitary and norm preserving.
\par We next consider the canonical transformation induced on ${\bf x}_s(\eta)$ 
by the functional $\hat{\bf G}_{\tilde f}$ defined in $(3.12)$.
Introducing a parameter $\hat \gamma$ along the motion generated by 
$\hat{\bf G}_{\tilde f}$, we have in this case by $(2.10a)$,
$$d{\bf x}_s(\eta)=\{ {\bf x}_s(\eta) , \hat{\bf G}_{\tilde f} \}
d \hat \gamma .\eqno(3.22)$$
Comparing $(2.10b)$ with $(3.13)$ and $(3.22)$ gives 
$$ \eqalign{{dq_s \over d\hat{\gamma}} &= \epsilon_s (-1)^F 
[(-1)^F \tilde f, q_s]
=\epsilon_s \tilde f q_s -q_s \tilde f,\cr
{dp_s \over d\hat{\gamma}} &= (-1)^F [(-1)^F \tilde f, p_s ]
=\tilde f p_s - \epsilon_s p_s \tilde f.\cr}\eqno(3.23)$$
For the bosonic sector, $(3.23)$ can be rewritten as 
$$\eqalign{
{d q_s \over d \hat{\gamma} }=&[\tilde f, q_s], \cr
{d p_s \over d \hat{\gamma} }=&[\tilde f, p_s],\cr 
}\eqno(3.24)$$
and can be integrated as a unitary
transformation for both $q_s$ and $p_s$, 
 $$ x_s(\hat{\gamma}) = e^{\tilde f \hat{\gamma}} x_s(0)
e^{-\tilde f \hat{\gamma}}. \eqno(3.25)$$

\par For the fermionic sector, however, the grading index $(-1)^F$ anticommutes 
with $q_s$ and $p_s$ and $\epsilon_s=-1$; consequently, the differential 
equations $(3.23)$ in this case take the form
$$ \eqalign{
{dq_s \over d\hat{\gamma}}&= -\{ \tilde f, q_s \} , \cr
{dp_s \over d\hat{\gamma}} &= \{ \tilde f,  p_s  \} ,\cr
} \eqno(3.26)$$
and involve {\it anticommutators} with the operator $\tilde f$, i.e., 
a graded action.
We note, however, that the total trace Lagrangians for which 
$\hat{\tilde C}$ is conserved are ones in which the fermion fields appear as 
bosonic bilinears of the form $p_r q_s$; for these bilinears,  
and for the reverse ordered bosonic bilinears $q_s p_r$, 
we find from $(3.26)$  that 
$$\eqalign{
{d (p_rq_s) \over d \hat{\gamma}}&=[\tilde f,p_r q_s], \cr
{d (q_sp_r)\over d \hat{\gamma}}&=-[\tilde f, q_sp_r]. \cr
}\eqno(3.27)$$
The solution of these differential equations is the unitary group action 
$$\eqalign{
(p_rq_s)(\hat{\gamma})&=e^{\tilde f \hat{\gamma}}(p_rq_s)(0)
e^{-\tilde f \hat{\gamma}}, \cr
(q_sp_r)(\hat{\gamma})&=e^{-\tilde f \hat{\gamma}} (q_sp_r)(0)
e^{\tilde f \hat{\gamma}}, \cr
}\eqno(3.28)$$
which preserves the supremum operator norm of the bilinears $p_rq_s$ and 
$q_sp_r$.  However, it is easy to see that for fermionic operators, 
the supremum 
operator norm of is not preserved by the evolution of $(3.26)$.
\par  Finally, it is also useful to define parameters $\gamma_{\pm}$ along 
the flows generated by ${\bf G}_{\pm \tilde f}$ according to 
$$d{\bf x}_s(\eta)=\{ {\bf x}_s(\eta), {\bf G}_{\pm \tilde f} \}
d \gamma_{\pm}, \eqno(3.29)$$
so that 
$${d x_s \over d \gamma_{\pm} }={1 \over 2} \left( {d x_s \over d \gamma }
\pm {d x_s \over d \hat{\gamma} } \right) .  \eqno(3.30)$$
Then taking sums and differences of $(3.20)$ and $(3.24),~(3.26)$ we 
find that for bosons (with $x_s$ either $q_s$ or $p_s$), 
$$\eqalign{
{d x_s \over d \gamma_+ }=&[\tilde f, x_s] , \cr
{d x_s \over d \gamma_- }=&0, \cr
}\eqno(3.31)$$
which integrate to 
$$\eqalign{
x_s(\gamma_+)=&e^{\tilde f \gamma} x_s(0) e^{-\tilde f \gamma}  , \cr
x_s(\gamma_-)=&x_s(0) .\cr
}\eqno(3.32)$$
Similarly, for fermions we find that 
$$\eqalign{
{d q_s \over d \gamma_+}=&-q_s\tilde f~,~~{d p_s \over d \gamma_+}=\tilde f~p_s , \cr  
{d q_s \over d \gamma_-}=&\tilde f~q_s~,~~~~~{d p_s \over d \gamma_-}=-p_s\tilde f , \cr  
}\eqno(3.33)$$
which integrate to
$$\eqalign{
q_s(\gamma_+)=&q_s(0)e^{-\tilde f \gamma_+}~,~~
p_s(\gamma_+)=e^{\tilde f \gamma_+} p_s(0), \cr
q_s(\gamma_-)=&e^{\tilde f \gamma_-}q_s(0)~,~~
~~p_s(\gamma_-)=p_s(0)e^{-\tilde f \gamma_-} . \cr
}\eqno(3.34)$$
This identifies ${\bf G}_{\pm \tilde f}$ as the generators of the one-sided 
unitary transformations acting on the fermions which are discussed in 
refs. 2 and 3.

\bigskip
\noindent
{\bf 4. The microcanonical and canonical ensembles.}
\smallskip
\par Introducing a complete set of states $\{ \vert n \rangle \}$ in
the underlying Hilbert space, the phase space operators are completely
characterized by their matrix elements $\langle m \vert x_r \vert n
\rangle \equiv (x_r)_{mn}$, which have the form
$$ (x_r)_{mn} = \sum_A (x_r)^A_{mn}e_A , \eqno(4.1)$$
where $A$ takes the values $0,\,1$ for complex Hilbert space,
$0,\,1,\,2,\,3$ for quaternion Hilbert space (technically, a Hilbert module), 
and just the one value $0$
for real Hilbert space, and the $e_A$ are the associated hypercomplex
units (unity, complex, or quaternionic units$^2$).  We restrict
ourselves here to these three cases. The phase space
measure is then defined
as
$$  d\mu = \prod_{r,m,n,A} d(x_r)^A_{mn}, \eqno(4.2)$$
where redundant factors are omitted according to adjointness
conditions.  The measure defined in this way is invariant under
canonical transformations induced by the generalized Poisson bracket.$^1$
\par We then define the microcanonical ensemble in terms of the set of
states in the underlying Hilbert space which satisfy $\delta$-function
constraints on the values of the two total trace functionals ${\bf H}~,~~
\hat {\bf H}$ and the 
matrix elements of the two conserved
operator quantities $\tilde C~,~~\hat{\tilde C}$ 
discussed in Section 1.  The volume of the               
corresponding submanifold in phase space is given by [see ref. 11 for
a somewhat more detailed discussion]
$$ \eqalign{\Gamma( E, {\hat E}, {\tilde\nu}, {\hat {\tilde\nu}}) &= \int d\mu\,\delta( E-{\bf
H})\,\delta({\hat E}- {\hat{\bf H}})\cr
& \prod_{n\leq m,A}\, \delta(\nu_{nm}^A
- \langle n \vert (-1)^F {\tilde C} \vert m \rangle^A)\, \delta({\hat 
\nu}_{nm}^A - \langle n \vert {\hat{\tilde C}} \vert m
\rangle^A ),\cr}  \eqno(4.3)$$
where we have used the abbreviations ${\tilde\nu} \equiv \{\nu_{nm}^A \}$ and
${\hat {\tilde\nu}} \equiv \{{\hat \nu}_{nm}^A \}$ for the parameters in the
arguments on the left hand side. The factor $(-1)^F$ in the term with
${\tilde C}$ is not essential, but convenient in obtaining the 
precise form given in ref. 1 for the canonical distribution.
The entropy associated with this ensemble is  given by
$$ S_{mic}(E, {\hat E}, {\tilde\nu}, {\hat {\tilde\nu}}) = \log \, \Gamma(E, {\hat E}, {\tilde\nu},
{\hat {\tilde\nu}}).\eqno(4.4)$$

\par The operators ${\tilde C}$ and ${\hat{\tilde C}}$ are defined in
terms of sums over degrees of freedom.  In the context of the
application to quantum field theory, the enumeration of degrees of
freedom includes continuous parameters, corresponding to the measure
space of the fields. These operators may therefore be decomposed into
parts within a certain (large) region of the measure space, which we
denote as $b$, corresponding to what we shall consider as a {\it
bath}, in the sense of statistical mechanics, and within another (small) 
part of the measure space, which we denote as $s$, corresponding to 
what we shall consider as a {\it subsystem}. We shall assume that the 
functionals ${\bf H}$ and ${\hat{\bf H}}$ may also be decomposed
additively into parts associated with $b$ and $s$; this assumption is
equivalent to the presence of interactions in the Hamiltonian or
Lagrangian operators
which are reasonably localized in the measure space of the fields (the
difference in structure between the Lagrangian and Hamiltonian
consists of operators that are explicitly additive), so that the
errors in assuming additivity are of the nature of ``surface terms''.
The constraint parameters may then be considered to be approximately
additive as well, and we may rewrite the microcanonical ensemble as

$$ \Gamma(E, {\hat E}, {\tilde\nu}, {\hat {\tilde\nu}}) = \int dE_s d{\hat E}_s 
(d\nu^s)(d{\hat
\nu}^s) \, \Gamma_b (E-E_s, {\hat E}-{\hat E}_s, {\tilde\nu} -
{\tilde\nu}_s,
 {\hat{\tilde
\nu}} - {\hat {\tilde\nu}}_s)\, \Gamma_s(E_s, {\hat E}_s, {\tilde\nu}_s, {\hat
{\tilde\nu}}_s). \eqno(4.5)$$ 
 \par We now assume that the integrand in $(4.5)$ has a maximum which
 dominates the integral when there is a large number of degrees of freedom.
 Let us, for brevity, define (we suppress the index $s$ in the following)
$$ \xi = \{ \xi_i \} \equiv \{E, {\hat E}, {\tilde\nu}, {\hat {\tilde\nu}}\},
 \eqno(4.6)$$ 
where the index $i$ refers to the elements of the set of variables, 
so that $(4.5)$ takes the form
$$\Gamma(\Xi)=\int d\xi \Gamma_b(\Xi-\xi) \Gamma_s(\xi), \eqno(4.7)$$
where $\Xi$ corresponds to the set of total properties for the whole 
ensemble.  
A necessary condition for an extremum in all of the variables at $\xi=\bar{\xi}$ 
is then (for every $i$)
$${\partial \over \partial \xi_i} [\Gamma_b(\Xi-\xi) \Gamma_s(\xi)]|_{\bar \xi}=0, \eqno(4.8)$$
which implies that 
$$ { 1 \over \Gamma_s (\xi)} {\partial \Gamma_s  \over
\partial \xi_i}  (\xi) |_{\bar \xi} = { 1 \over \Gamma_b(\Xi - \xi)}
{\partial  \Gamma_b\over \partial \Xi_i }(\Xi -
\xi)|_{\bar \xi}.
\eqno(4.9)$$
The logarithmic
derivatives in $(4.9)$ define a set of quantities analogous to the
(reciprocal) temperature of the usual statistical mechanics, i.e.,
equilibrium-fixing Lagrange parameters common to the bath and the
subsystem.  We write these separately as
$$\eqalign{ \tau &= {\partial \over \partial E } \log \Gamma_s
(\xi)|_{\bar \xi} \cr
            {\hat \tau} &= {\partial \over \partial {\hat E}} \log
\Gamma_s(\xi)|_{\bar \xi} \cr
 \lambda_{nm}^A &= -{\partial \over \partial\nu_{nm}^A } \log
\Gamma_s(\xi)|_{\bar \xi} \cr
{\hat \lambda}_{nm}^A &= -{\partial \over \partial {\hat \nu}_{nm}^A}\log
\Gamma_s(\xi)|_{\bar \xi}. \cr} \eqno(4.10)$$
According to the definition of entropy $(4.4)$, the bath phase space
volume
is given by
$$\eqalign{ \Gamma_b(\Xi-\xi_s) &= e^{S_b(\Xi -\xi_s)} \cr
                              &\cong e^{S_b(\Xi)} \exp \{-\sum_i \xi_{i,s}
{\partial S_b \over \partial \Xi_i } (\Xi)\},\cr} \eqno(4.11)$$
Neglecting the small shift in argument $\Xi \rightarrow \Xi - \xi_s$, it
follows from  $(4.7)-(4.10)$ that
$$  \Gamma_b (\Xi-\xi_s) \cong e^{S_b(\Xi)} \exp\{-\tau E - {\hat
\tau}{\hat E} + \sum_{n \leq m,A} (\nu_{nm}^A \lambda_{nm}^A +
{\hat \nu}_{nm}^A {\hat \lambda}_{nm}^A)\}. \eqno(4.12)$$
\par We now return to $(4.5)$, replacing the phase space volume of the
 bath,  $\Gamma_b$, by the approximate form $(4.12)$, and the
subsystem phase space volume $\Gamma_s$ by the phase space integral
over the constraint $\delta$-functions.
\par Carrying out the integrals over the parameters, the
$\delta$-functions imply the replacement of the parameters
$E$, ${\hat E}$, $\nu_{nm}^A$, ${\hat \nu}_{nm}^A$ in the
exponent by the corresponding phase space quantities.  Using the
anti-self-adjoint properties of ${\tilde C}$ and ${\hat{\tilde C}}$,
 and defining the operator ${\tilde \lambda}$ for which the matrix elements
are $\langle n \vert {\tilde \lambda} \vert m \rangle ={ 1 \over 2}
\lambda_{nm},~~n\neq m,$, and $\langle n \vert {\tilde \lambda} \vert
n \rangle =
\lambda_{nn},$ we see that the first term in the sum in the exponent
in $(4.12)$ is $-{\bf Tr} {\tilde \lambda}{\tilde C}$.  A similar
result holds for the last term of the sum (in this case, since we did
not insert the factor $(-1)^F$, we obtain the ${\hat{\bf Tr}}$ functional). 
The volume in phase space is then 
$$ \Gamma (\Xi) = e^{S_b(\Xi)} \int d\mu_s \exp - \{ \tau {\bf H}_s +
{\hat \tau} {\hat {\bf H}}_s + {\bf Tr} {\tilde \lambda} {\tilde
C}_s + {\hat {\bf Tr}} {\hat {\tilde  \lambda}} {\hat {\tilde C}}_s
\}, \eqno(4.13)$$
 so that the normalized canonical distribution function is given by 
 $$ \rho = Z^{-1} \exp- \{ \tau {\bf H} +
{\hat \tau} {\hat {\bf H}} + {\bf Tr} {\tilde \lambda} {\tilde
C} + {\hat {\bf Tr}} {\hat {\tilde  \lambda}} {\hat {\tilde C}}
\}, \eqno(4.14)$$
where $$ Z = \int d\mu\, \exp- \{ \tau {\bf H} +
{\hat \tau} {\hat {\bf H}} + {\bf Tr} {\tilde \lambda} {\tilde
C} + {\hat {\bf Tr}} {\hat {\tilde  \lambda}} {\hat {\tilde C}}
\}. \eqno(4.15)$$
\par This formula coincides with that obtained by Adler and
Millard.$^1$ Note that the operators ${\tilde \lambda}$ and
${\hat{\tilde \lambda}}$ appear as an infinite set of inverse
``temperatures'', i.e., equilibrium Lagrange parameters associated both with
the bath and the subsystem, corresponding to the conserved matrix elements of
$(-1)^F {\tilde C}$ and ${\hat{\tilde C}}$. 
\par If ${\hat{\tilde \lambda}}$ is a function of ${\tilde \lambda}$,
then the ensemble average of an operator $\cal O$, which we denote as
$\langle {\cal O}\rangle_{AV}$, is a function of ${\tilde \lambda}$
(and other scalar parameters), and hence commutes with it.  If we
diagonalize $\langle {\tilde C} \rangle_{AV}$ to the form $i_{eff}D$,
and assume that $D$ is totally degenerate$^1$, and equal to $\hbar$,
then the analog of the
equipartion (Ward) identities gives$^1$
$$ 0 = \langle \hbar ({\dot x}_r)_{eff} - i_{eff} [ H_{eff},
(x_r)_{eff}] \rangle_{AV}, $$
where we write the subscript $eff$ to indicate the part of the
operator that commutes with $i_{eff}$.  All of the complex canonical
structure is then reproduced in the ensemble average.  A second set of
Ward identites$^1$ implies that for fermionic $q_r,\, p_r$, 
$$\langle {\tilde H}_{eff} q_r \rangle_{AV} = \langle p_r {\tilde
H}_{eff} \rangle_{AV} = 0; $$
from this, one sees that $\langle\, {}\,\rangle_{AV}$ appears to have
properties of the Wightmann vacuum.
\par We finally remark that the microcanonical entropy defined in
$(4.4)$ provides the Jacobian of the transformation from the
integration over the measure of $\cal S$ in $(3.15)$ to an integral
over the parameters defining the microcanonical shells. To see this,
we rewrite $(4.15)$ as 
$$ \eqalign {Z &= \int d\mu dE d{\hat E} (d\nu)(d\hat{\nu})\delta( E-{\bf
H})\,\delta({\hat E}- {\hat{\bf H}})\,\cr
 &\times \prod_{n\leq m,A}\,
 \delta(\nu_{nm}^A
- \langle n \vert (-1)^F {\tilde C} \vert m \rangle^A)\, \delta({\hat 
{\nu}}_{nm}^A - \langle n \vert {\hat{\tilde C}} \vert m 
\rangle^A )\cr
& \times  \exp -\{ \tau E +
{\hat \tau} {\hat E} + {\bf Tr} {\tilde \lambda} {\tilde\nu} 
+ {\hat {\bf Tr}} {\hat {\tilde  \lambda}} {\hat {\tilde\nu}}
\}, \cr} \eqno(4.16)$$
where we have defined the anti-self-adjoint parametric operators
${\tilde\nu}$ and ${\hat{\tilde\nu}}$ by
$$\eqalign{ \nu_{nm}^A &= \langle n \vert (-1)^F {\tilde \nu} \vert m
\rangle, \cr
{\hat \nu}_{nm}^A &=  \langle n \vert {\hat {\tilde \nu}}\vert m
\rangle^A .\cr } \eqno(4.17)$$
 The phase space integration over the $\delta$-function
factors reproduces the volume of the microcanonical shell associated
with these parameters, i.e, the exponential of the microcanonical
entropy, so that the partition function can be written as
$$ Z = \int dE d{\hat E} (d{\nu})(d{\hat {\nu}}) e^{S_{mic}(E,
{\hat E}, {\nu}, {\hat{\nu}})} \exp-\{\tau  E +
{\hat \tau} \hat E + {\bf Tr} {\tilde \lambda} {\tilde\nu} + 
{\hat {\bf Tr}} {\hat {\tilde  \lambda}} {\hat {\tilde\nu}}
\}. \eqno(4.18)$$
\par We now turn to  study the stability of the canonical ensemble
as associated with the dominant contribution to the microcanonical
phase space volume. To this end, we formally define the free
energy $A$ as
the negative of the logarithm of the partition
function,\footnote{$^\sharp$}{The conventional symbol for the free energy
 should not be confused with the hypercomplex index $A$.} 
$$ Z \equiv e^{-A(\tau, {\hat \tau},{\tilde \lambda}, {\hat{\tilde
\lambda}})}, \eqno(4.19)$$
so that $(4.15)$ can be written as
$$ 1 = \int d\mu e^{A(\tau, {\hat \tau},{\tilde \lambda}, {\hat{\tilde
\lambda}})}\exp{- \{ \tau {\bf H} +
{\hat \tau} {\hat {\bf H}} + {\bf Tr} {\tilde \lambda} {\tilde
C} + {\hat {\bf Tr}} {\hat {\tilde  \lambda}} {\hat {\tilde C}}
\}}. \eqno(4.20)$$
Let us now define the set of variables
$$ \{\chi_i\} = \{ \tau, {\hat \tau}, -\lambda_{nm}^A,
-{\hat\lambda}_{nm}^A \}. \eqno(4.21)$$
Differentiating $(4.20)$ with respect to each of these, one finds (as
in Eqs. $(49)$ of ref .1) the
relations$^{11}$ 
$$ {\partial A \over \partial \chi_i } = \xi_i,  \eqno(4.22)$$
where we have associated the average values of the dynamical variables
 ${\bf H}$, ${\hat{\bf H}}$,
 $C_{nm}^A$ and ${\hat C}_{nm}^A$  with $E$, ${\hat E}$, 
$ \nu_{nm}^A$ and
${\hat \nu}_{nm}^A $.
\par We now consider the identity
$$ 0 = \int d\mu\, ({\bf H} - \langle {\bf H} \rangle_{AV})e^{A(\tau,
 {\hat \tau},{\tilde \lambda}, {\hat{\tilde
\lambda}})}
\exp{- \{\tau {\bf H} +
{\hat \tau} {\hat {\bf H}} + {\bf Tr} {\tilde \lambda} {\tilde
C} + {\hat {\bf Tr}} {\hat {\tilde  \lambda}} {\hat {\tilde C}}
\}}. \eqno(4.23)$$
Differentiating with respect to $\tau$, one finds
$$ \eqalign{0 = \int d\mu\,\bigl( {\partial A \over \partial \tau} -
{\bf H} \bigr)&({\bf H} - \langle {\bf H} \rangle_{AV})e^{A(\tau,
 {\hat \tau},{\tilde \lambda}, {\hat{\tilde
\lambda}})}\cr
&\times \exp{- \{ \tau {\bf H} +
{\hat \tau} {\hat {\bf H}} + {\bf Tr} {\tilde \lambda} {\tilde
C} + {\hat {\bf Tr}} {\hat {\tilde  \lambda}} {\hat {\tilde C}}
\}} - {\partial \langle {\bf H} \rangle_{AV}
 \over \partial \tau} , \cr}                    \eqno(4.24)$$
so that, from $(4.22)$, we find that (as in ref. 1)
$$ \langle ({\bf H} - \langle {\bf H} \rangle_{AV})^2 \rangle_{AV} =
-{\partial \langle {\bf H} \rangle_{AV} \over \partial \tau} = -
{\partial^2 A \over \partial \tau^2} \geq 0. \eqno(4.25)$$
\par In fact, applying this argument to all of the dynamical
quantities (including cross terms), one can  show that $A$ is a
locally convex function.$^{11}$  The
Taylor series expansion of $A(\xi + \delta \xi)$, up to second order,
is, with the help of the relations obtained in this way,  given by 
$$\eqalign{
&A(\tau+\delta \tau,\hat \tau+ \delta \hat \tau, \tilde \lambda + \delta 
\tilde \lambda , \hat {\tilde \lambda} + \delta \hat{\tilde \lambda})\cr
=&A(\tau,\hat \tau,\tilde \lambda, \hat{\tilde \lambda}) +\delta \tau 
\langle {\bf H} \rangle_{AV} +\delta \hat{\tau} \langle \hat{\bf H} 
\rangle_{AV}-\sum_{n\leq m, A}(\delta \lambda_{nm}^A \langle C_{nm}^A 
\rangle_{AV} + \delta \hat{\lambda}_{nm}^A \langle \hat C_{nm}^A
\rangle_{AV}) \cr
-&{1\over2} \langle [ \delta \tau({\bf H} -\langle {\bf H}\rangle_{AV})
+\delta \hat{\tau} (\hat{\bf H}-\langle \hat{\bf H} \rangle_{AV}) \cr
-&\sum_{m\leq n, A} \delta \lambda_{nm}^A (C_{nm}^A-\langle C_{nm}^A 
\rangle_{AV}) + \delta \hat{\lambda}_{nm}^A (\hat C_{nm}^A - \langle 
\hat C_{nm}^A \rangle_{AV} ) ]^2 \rangle_{AV}; \cr
}\eqno(4.26)$$
the uniform negative sign of the quadratic term in the expansion indicates 
that $A$ is a locally convex function, and shows that the matrix of 
second derivatives of $A$ is negative semidefinite.
\par We now turn to the alternative expression of $(4.18)$ for the partition
function, defined in terms of an integral over the parameters of a
sequence of microcanonical ensembles.  The existence of a
maximum in the integrand which dominates the integration assures the
stability of the canonical ensemble; we show that $(4.26)$ implies the 
self-consistency of our assumption of a maximum.  
\par The conditions for a maximum
of the integrand at $\xi = \bar \xi$ in $(4.18)$ are that there be a 
stationary point, i.e., 
that
 $$ \chi_i = {\partial S_{mic} \over \partial \xi_i}|_{\bar \xi},
\eqno(4.27)$$
 together with the requirement that the second derivative matrix 
$${\partial^2 S_{mic}\over \partial \xi_i \partial \xi_j} 
={\partial \chi_i \over \partial \xi_j} \eqno(4.28)$$
should be positive definite.  
\par The matrix inverse to the right hand side of $(4.28)$ is given by
$$ {\partial \xi_j \over \partial \chi_i } = {\partial^2 A\over
\partial \chi_i \partial \chi_j}, \eqno(4.29)$$
which we have shown to be a negative semidefinite matrix.  This in turn 
implies that the matrix on the right hand side of $(4.28)$ is 
negative definite, 
giving the condition needed to assure that the stationary point  in
$(4.27)$ is indeed a (local) maximum.
\par Assuming this maximum dominates the integration, then the logarithm
of the integral in $(4.18)$ (up to an additive term which is
relatively small for a large number of degrees of freedom) may be
approximated by
$$ A \cong \tau E + {\hat \tau} {\hat E} + {\bf Tr}{\tilde
\lambda}{\tilde C} + {\hat {\bf Tr}} {\hat{\tilde \lambda}}
{\hat{\tilde C}} - S_{mic}( E,\, {\hat E}, {\tilde C}, \,  {\hat{\tilde
C}}), \eqno(4.30)$$
where the arguments are at the extremal values, giving the analog of the   
standard thermodynamical result  $A = E - TS$ for the free energy.
\bigskip
\centerline{\bf Acknowledgments}
This work was supported in part by the Department of Energy under 
Grant \#DE-FG02-90ER40542.  

\centerline{\bf References}
\item{1.} S.L. Adler and A.C. Millard, ``Generalized Quantum Dynamics 
as Pre-Quantum Mechanics'', Nuc. Phys. {\bf B}, in press.
\item{2.} S.L. Adler, Nuc. Phys. B {\bf 415} (1994) 195.
\item{3.} S.L. Adler, {\it Quaternionic Quantum Mechanics and 
Quantum Fields},  Oxford University Press, New York and Oxford, 1995.
\item{4.} S.L. Adler, Phys. Rev. D {\bf 21}, 2903 (1980), and
references therein.
\item{5.} E.C.G. Stueckelberg, Helv. Phys. Acta. {\bf 33} (1960) 727;   
{\bf 34} (1961)  621, 675; {\bf 35} (1962) 673. 
\item{6.} D. Finkelstein, J.M. Jauch, S. Schiminovich, and D. Speiser,
J. Math. Phys. {\bf 3} (1962) 207; {\bf 4} (1963) 788. 
\item{7.} L.P. Horwitz and L.C. Biedenharn, Ann. Phys. {\bf 157} (1984) 432. 
\item{8.} C. Piron, {\it Foundations of Quantum Physics}, W.A. Benjamin,
Reading, MA, 1976.
\item{9.} S.L. Adler, G.V. Bhanot, and J.D. Weckel, J. Math. Phys.
{\bf 35} (1994) 531.
\item{10.} S.L. Adler and Y.-S. Wu, Phys. Rev. {\bf D49} (1994) 6705.
\item{11.} S.L. Adler and L.P. Horwitz, ``Microcanonical Ensemble and
Algebra of Conserved Generators for Generalized Quantum Dynamics,''
Institute for Advanced Study Preprint IASSNS 96/36, submitted for
publication to J. Math. Phys.

\end 
\bye